# RECENT PROGRESS ON A MANIFOLD DAMPED AND DETUNED STRUCTURE FOR CLIC

V.F. Khan[†*], A. D'Elia[†*‡], A. Grudiev[‡], R.M. Jones[†*], W. Wuensch[‡], R. Zennaro[‡]
†School of Physics and Astronomy, The University of Manchester, Manchester, U.K.
[*] The Cockcroft Institute of Accelerator Science and Technology, Daresbury, U.K.
[‡] CERN, Geneva, Switzerland.

*Abstract*

A damped detuned structure for the main X-band linacs of CLIC is being investigated as an alternative design to the present baseline heavily damped structure [1]. In our earlier designs [2-6] we studied detuned structures, operating at 11.994 GHz, with a range of dipole bandwidths in order to ensure the structure satisfies beam dynamics and rf breakdown constraints [1,7]. Here we report on the development of a damped and detuned structure which satisfies both constraints. Preparations for high power testing of the structure are also discussed.

## INTRODUCTION

The baseline design for the CLIC main linacs operates at an accelerating gradient of 100 MV/m [1,7]. This imposes strong constraints on the e.m. fields which can be tolerated within the structure. It is important to ensure the surface e. m. fields are minimised in order to prevent electrical breakdown. In addition, wakefield excitation must be suppressed [8] in order to contain emittance dilution. Surface fields are minimised by suitably designing the cavity shape and by ensuring defects in the surface morphology are minimised by surface processing methods [9]. The wakefield is suppressed, either by providing heavy damping (as is done in the baseline design [1,7] in which the dipole Q~10) or by a combination of moderate damping and strong detuning [10,11]. This alternative method is followed herein.

During the course of our study we have investigated three main approaches. In all cases we require frequencies of 8 structures to be interleaved, and hence we effectively have 192 cells (24 cells per structure). The phase advance per cell is $2\pi/3$ radians. The first approach focused on the beam dynamics. The fundamental mode frequency of 11.994 GHz results in a concentration of the kick factors [12] (which impart a transverse momentum kick to the beam) in the first dipole band and to a lesser degree in the third and sixth bands [13]. For this reason we focus on detuning these dipole bands by tapering the cavity irises and thicknesses. In this case the bandwidth of the first dipole band is required to be ~3.3 GHz in order to suppress the long range wakefield at each bunch (spaced from their neighbours by 6 rf cycles or ~ 0.5 ns) [2]. However, this results in unacceptably high surface e.m. fields. A modified design, tied closely to the CLIC baseline end cells, resulted in a bandwidth of ~ 1GHz [3, 4]. However this requires each bunch to be positioned at the zero crossing in the dipole wakefields and, although not extensively investigated, may have imposed stringent fabrication tolerances. The third design entails modifying the bunch spacing to 8 rf cycles (0.67 ns) whilst ensuring the beam dynamics constraints imposed by the short and long range wakefields [7] are adhered to, together with sufficiently small surface fields. This approach, CLIC_DDS_C, is discussed in the following section. However, it results in an unacceptably high surface pulsed temperature rise ($\Delta T$ [9]) and consequently was modified with a design incorporating an elliptical outer wall, CLIC_DDS_E. This is discussed in a subsequent section, together with plans for a high power test of a single structure at CERN, in the penultimate section.

## RELAXED PARAMETERS : CLIC_DDS_C

Modifying the bunch spacing to 8 rf cycles allowed the transverse wakefield to be damped sufficiently once the structure is provided with a bandwidth of ~2.3 GHz. The dipole frequencies of 24 cells are detuned by enforcing an erf (error) variation of each iris with cell number. In this manner, the wakefield decays initially with a Gaussian roll -off. The bandwidth in terms of standard deviation $\sigma$ is, $\Delta f=3.6\sigma$, or 13.75% of the central frequency. However, as there are a finite number of cells, eventually the modes, which constitute the wakefield, recohere and the wakefield rises above the acceptable level. This would be the case if pure detuning was relied upon. However, in practice manifold damping is used in addition to detuning.

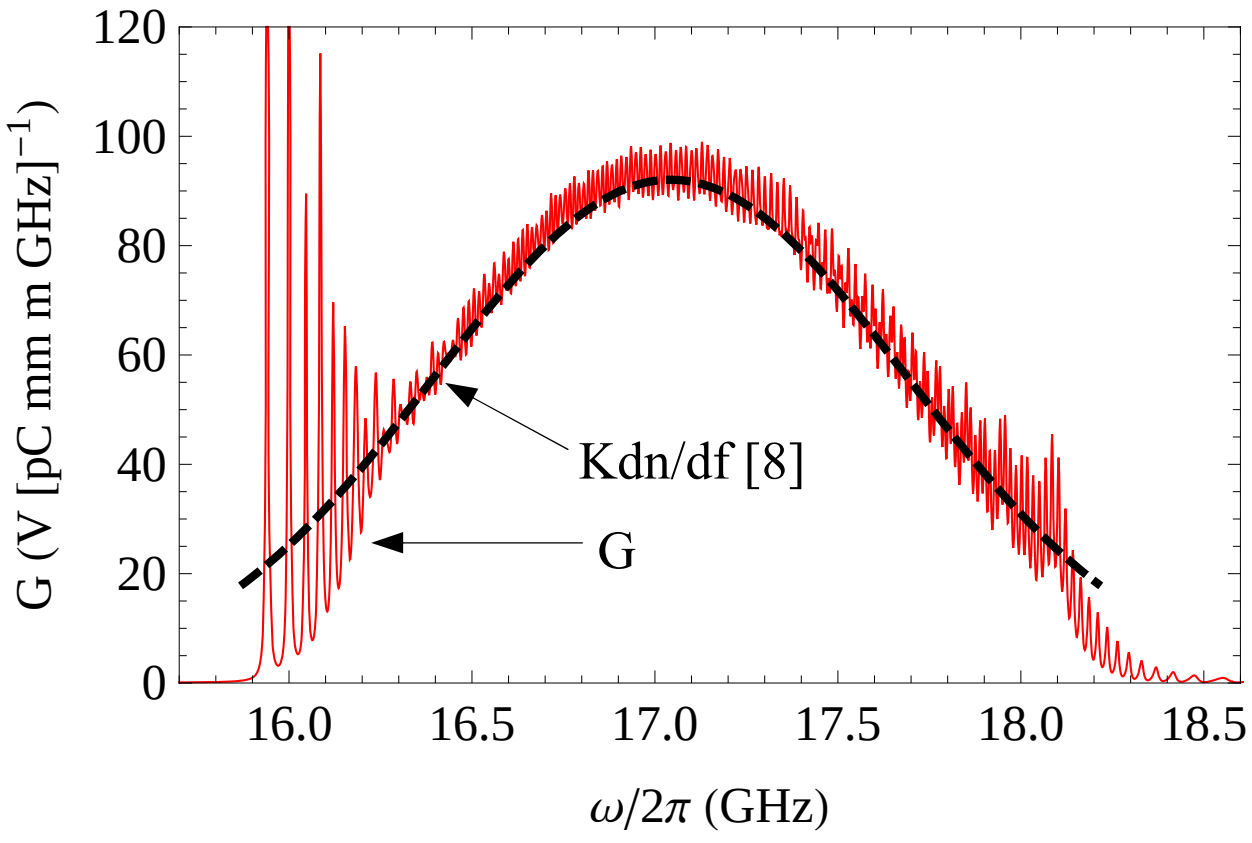

Figure 1: Spectral function (G) of an 8-fold interleaved CLIC_DDS_C

An indication of how well the modes are coupled to the attached manifold is provided by the spectral function [10], and this is illustrated in Fig. 1. This was obtained by simulating a select number of cells with HFSS in order to obtain parameters suitable for a circuit model to be applied. Taking the inverse Fourier transform allows the wakefield to be obtained and is displayed in Fig. 2 for a detuned structure both with (CLIC_DDS_C) and without (DS) manifold damping, but including Ohmic losses. It is clear that the beam dynamics constraints are well satisfied. However, extensive simulations, performed with HFSS and CST microwave studio, indicate that ΔT is too large for this structure. This led to a modification of the cavity outer wall, from its original circular shape to an elliptical shape. This is discussed in the next section.

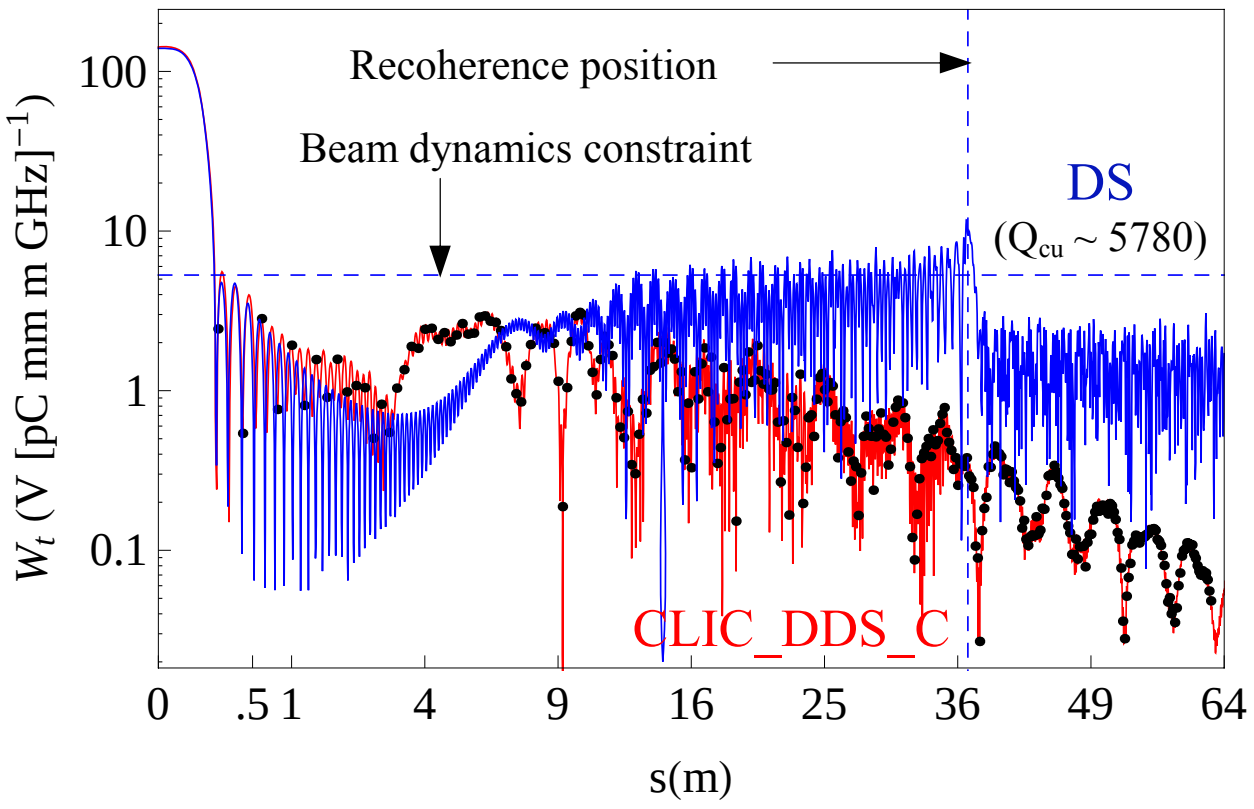

Figure 2: Envelope of the wakefield for an 8-fold interleaved CLIC_DDS_C

## ELLIPTICAL WALL : CLIC_DDS_E

The manifold is coupled to each accelerating cell through a slot in the cavity wall. However, this aperture enhances the magnetic field and hence increases ΔT. This can readily degrade the surface of the accelerator and in order to prevent this, ΔT must be minimised. We considered various elliptical geometries for the cavity wall in order to reduce the field in the region where the slot will be located. In this manner the overall rise in ΔT is reduced. The variation of magnetic field with ellipticity (ε) is illustrated in Fig. 3. We studied both convex and concave shapes. The variation of the surface magnetic field ($H_s$) along the border of the cavity is illustrated in Fig. 4. We chose ε = 1.2 in order to minimise the peak surface magnetic field. The overall pulsed temperature heating is then reduced from ~65 $^o$K to 51 $^o$K (~23 %).

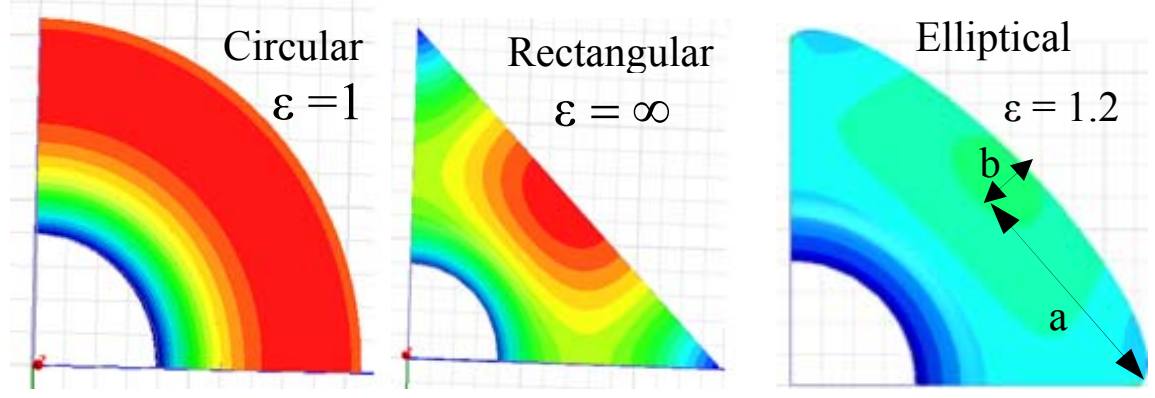

Figure 3: Magnetic field distribution in an undamped quarter symmetry cell for several ellipticities.

However, reducing the surface field of the main accelerating mode also diminishes the intensity of the dipole mode field in the vicinity of the slots. Consequently, the coupling of dipole modes to the manifold is reduced and the damping of the wakefield is degraded. Furthermore, practical mechanical engineering considerations, including the rounding of corners and edges, constrain the tip of the manifold to be located no more than 6.8 mm with respect to the centre of the cavity. The spectral function, which incorporates all of these effects, is illustrated in Fig. 5 for interleaving of the dipole frequencies of 8 accelerating structures.

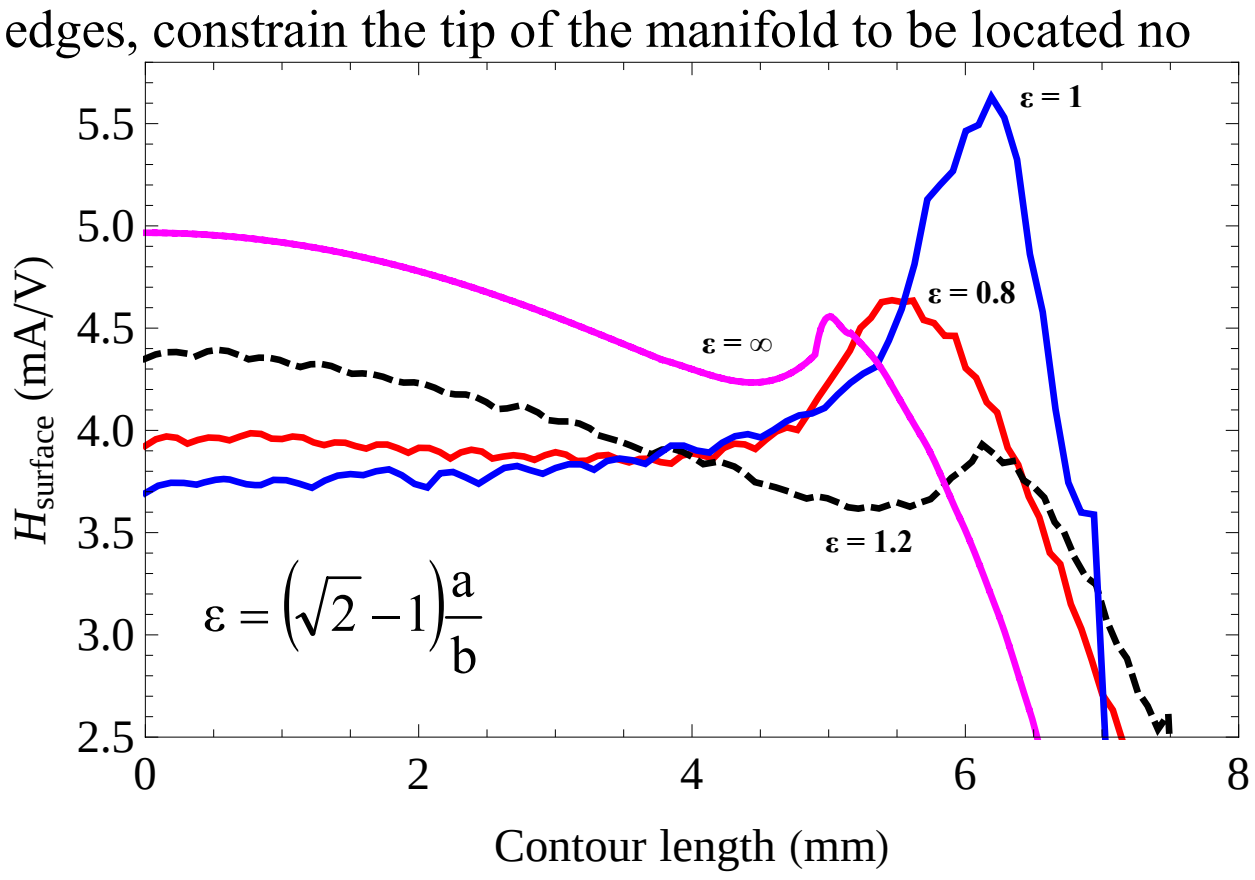

Figure 4: Normalised magnetic field on the cavity wall of a manifold damped cell for several geometries.

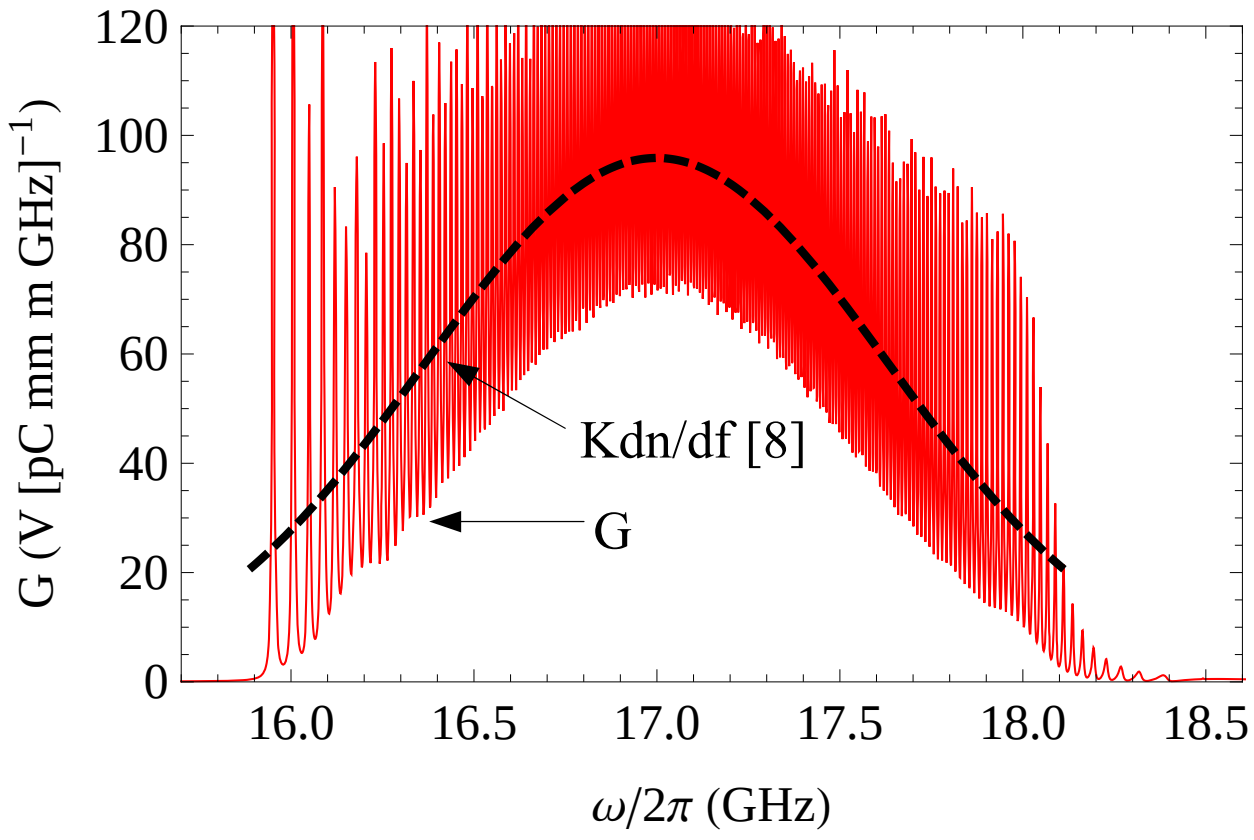

Figure 5: G of an 8-fold interleaved CLIC_DDS_E

The amplitude of the oscillations is significantly larger than that of Fig. 1. The surface magnetic field has been reduced at the expense of a non-optimal dipole coupling. Nonetheless, the wakefield, displayed in Fig. 6, and obtained from the inverse Fourier transform of the spectral function, is still acceptable from a beam dynamics perspective. The influence of non-optimal coupling is evident after ~10 m (~ 50 bunches) and although it is sufficiently suppressed to contain emittance dilution, fabrication tolerances are expected to be increased. In order to access the suitability of this design to sustain high fields, a structure will be fabricated and this is described in the following section.

## HIGH POWER TEST: CLIC_DDS_A

A structure will be fabricated and high power tested with an input power of 71 MW. For the sake of mechanical simplicity this structure has been equipped with a manifold of constant penetration into the cells (of 6.8 mm) and constant radius (2.1 mm). The latter was chosen to ensure the manifold mode is cut-off to the accelerating mode and hence has very little impact on the fundamental mode shunt impedance.

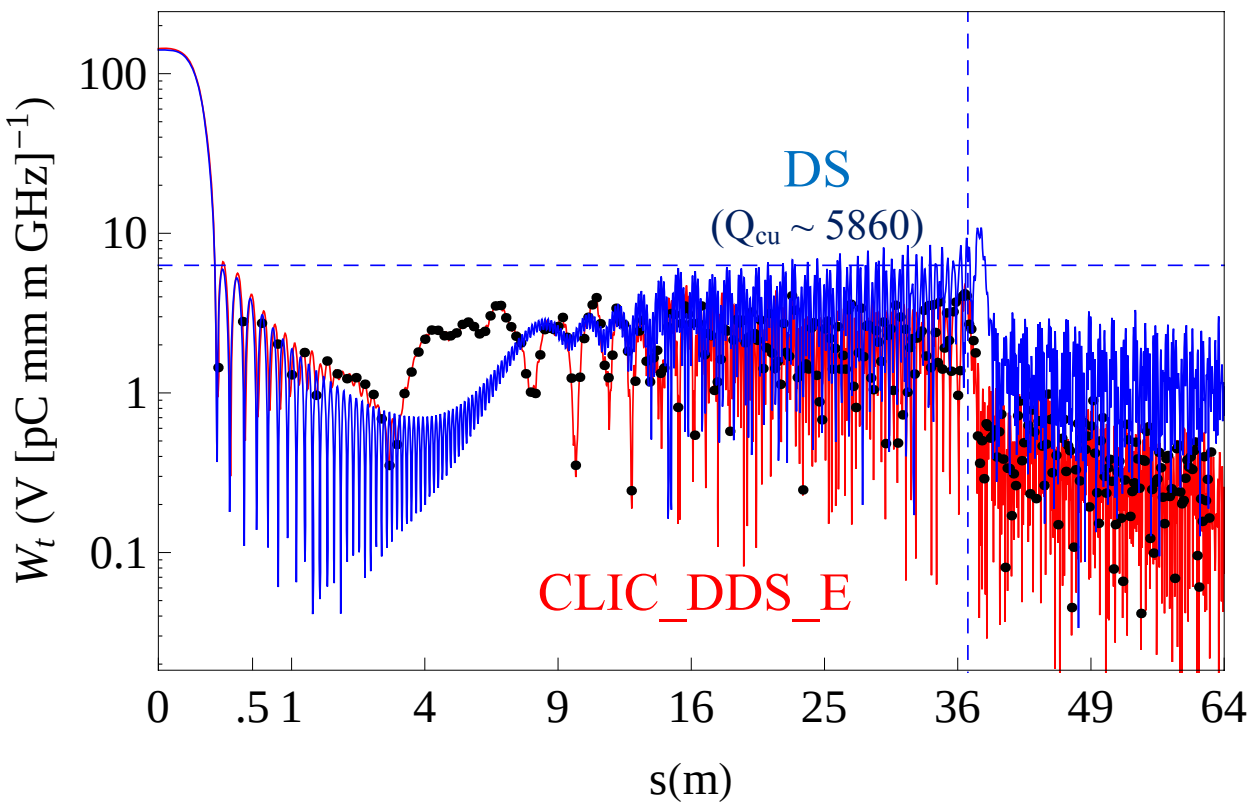

Figure 6: Envelope of wakefield for an 8-fold interleaved CLIC_DDS_E.

The iris radii of the cavities taper down from 4 mm to 2.5 mm and the thickness varies from 4 mm to 1.4 mm. In this manner the frequencies of the several bands are detuned. The cell geometry is presented in Fig. 7 in which two cells are displayed end-to-end.

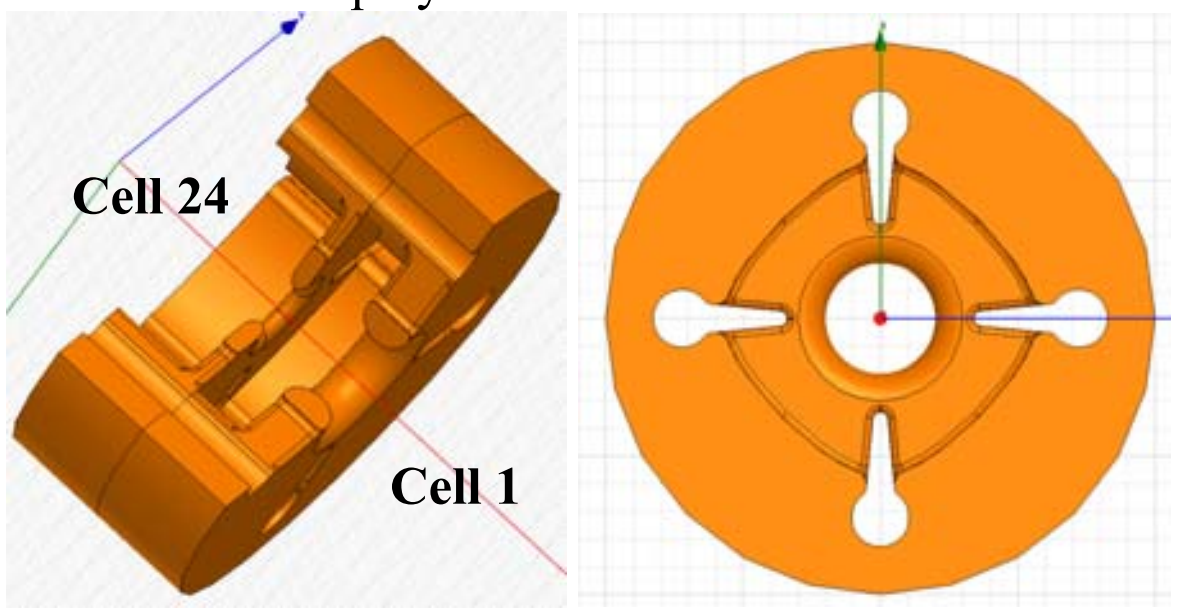

Figure 7: CAD representation of CLIC_DDS_A.

The surface fields for these have been simulated with HFSS and are shown in Fig. 8 for the cells which indicate maximum values. The rf properties of the structure are summarised in the Table. The overall rf properties, both with and without beam loading, are illustrated in Fig. 9. Apart from $S_c$ [14], their maximum values are below the requisite CLIC_G [1,7] baseline values.

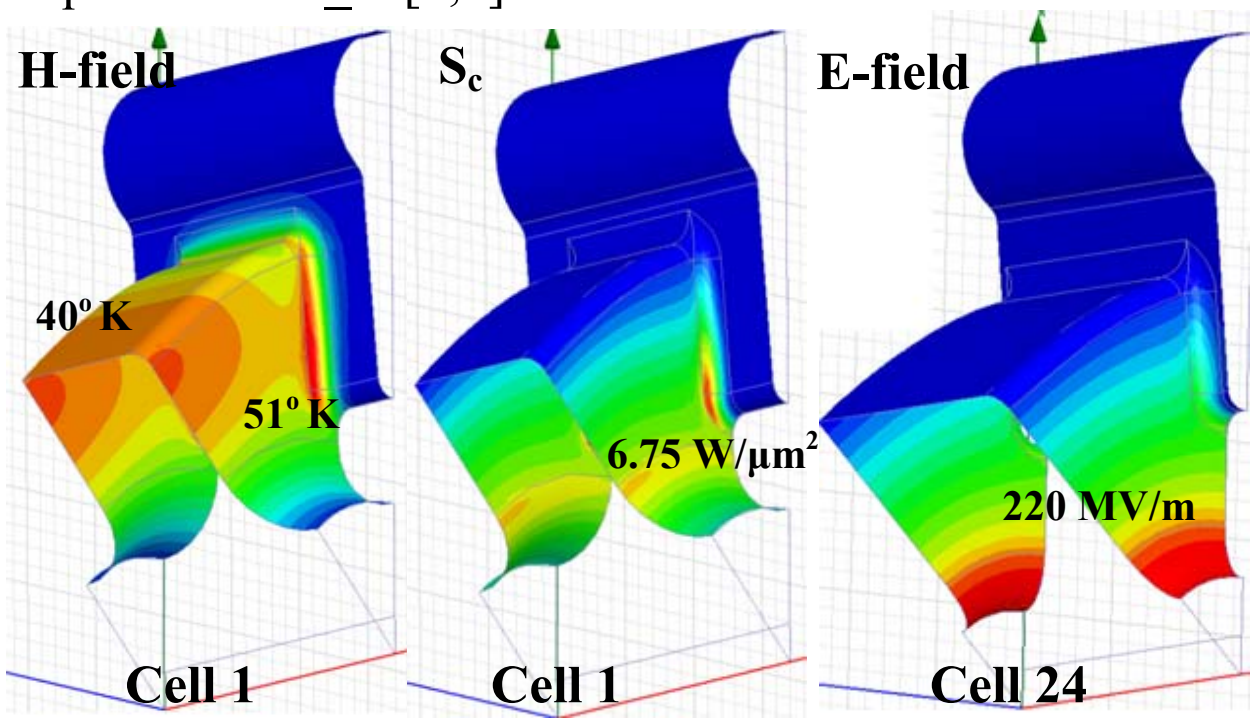

Figure 8: Surface fields in single cells of CLIC_DDS_A.

| **RF parameter** | **Unit** | **Value** |
|---|---|---|
| Fundamental Mode Q (In/Out) | - | 5020 / 6534 |
| Shunt impedance R' (In/Out) | MΩ/m | 51 / 118 |
| Group velocity $v_g/c$ (In/Out) | % | 2.07 / 1.0 |
| Bunch population ($N_b$) | $10^9$ | 4.2 |
| Peak input power ($P_{in}$) | MW | 71 |
| Pulse length (Total/ Flat top) | ns | 277 / 208 |
| Acc. grad. $E_{acc}^{max}$ (Load./Unload) | MV/m | 105 / 132 |
| Pulsed tempeature. rise $\left(\Delta T_{sur}^{max}\right)$ | $^o$K | 51 |
| Surface electric field $\left(E_{sur}^{max}\right)$ | MV/m | 220 |
| Modified Poynting vector $\left(S_c^{max}\right)$ | W/μm$^2$ | 6.75 |
| RF-beam efficiency (η) | % | 23.5 |

Table: RF parameters of a non interleaved 24 cell CLIC_DDS_A.

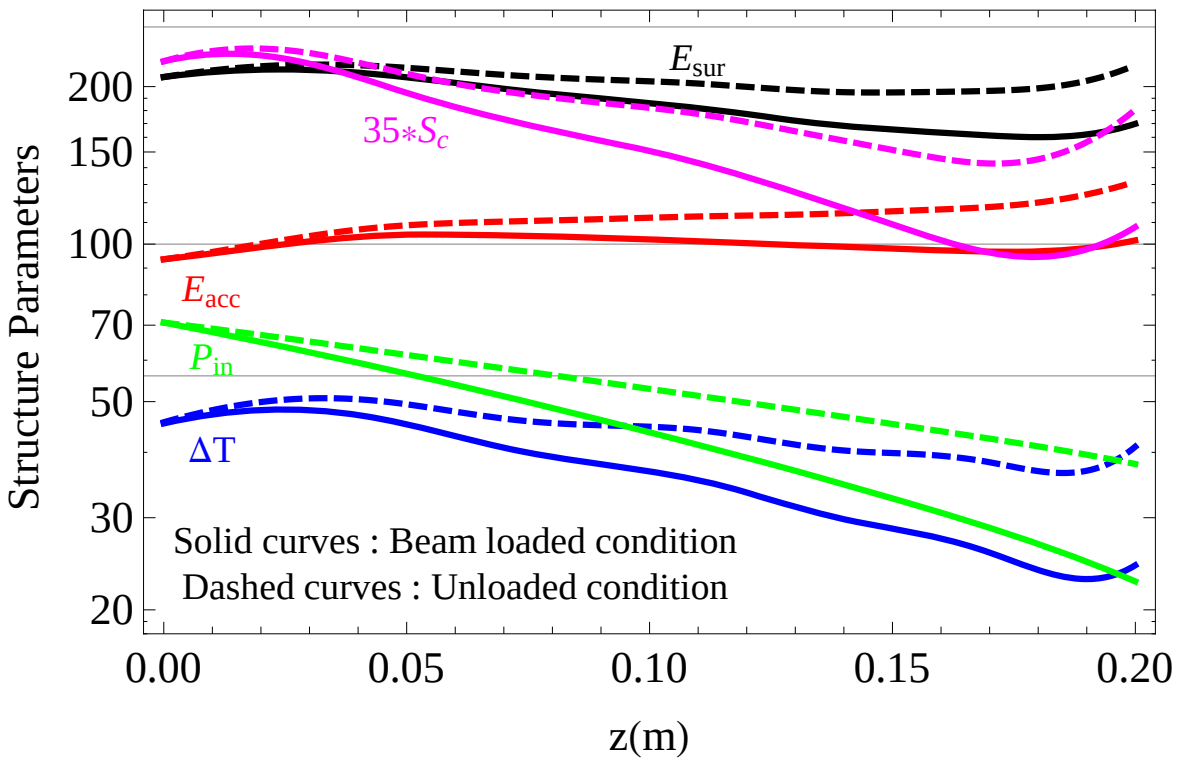

Figure 9: RF properties of a non-interleaved 24 cell structure: CLIC_DDS_A.

## FINAL REMARKS

Extensive investigations have allowed a test accelerator structure to be designed suitable for suppressing the long range wakefield in the CLIC main linacs. The surface e.m. fields have been minimised and the wakefield is well suppressed for an 8-fold interleaving of structures. A single structure will be fabricated and subjected to high power tests at CERN. Fabrication tolerances of these interleaved structures remain to be investigated.

## ACKNOWLEDGMENTS

V.F. Khan receives funding from the Cockcroft Institute. This research has received funding from the European Commission under the FP7 Research Infrastructure grant agreement no. 227579.